\documentclass[prl,twocolumn,amsmath]{revtex4-1}

\usepackage{graphicx}
\usepackage{amssymb,amsfonts,amsmath}

\def \e{\epsilon}    
   \def \d{\delta} 
    \def \l{\lambda}

\def \L{\Lambda} 

\def \del{\partial}    

\def \HF{\dfrac{1}{2}}  

\def \>{\rangle} 
\def \<{\langle} 
 
\def\be{\begin{equation}} 
\def\ee{\end{equation}} 
\def\longrightharpoonup{\relbar\joinrel\rightharpoonup}
\def\longleftharpoondown{\leftharpoondown\joinrel\relbar}

\def\longrightleftharpoons{
  \mathop{
    \vcenter{
      \hbox{
      \ooalign{
        \raise1pt\hbox{$\longrightharpoonup\joinrel$}\crcr
	  \lower1pt\hbox{$\longleftharpoondown\joinrel$}
	  }
      }
    }
  }
}

\newcommand \bea {\begin{eqnarray}} 
\newcommand \eea {\end{eqnarray}}

\begin{document}

\title{ Bayesian feature selection with strongly-regularizing priors maps to the Ising Model }

\author{ Charles K. Fisher and Pankaj Mehta}
\affiliation{Dept. of Physics, Boston University, Boston, MA 02215}

\date{\today}

\begin{abstract}
Identifying small subsets of features that are relevant for prediction and/or classification tasks is a central problem in machine learning and statistics. The feature selection task is especially important, and computationally difficult, for modern datasets where the number of features can be comparable to, or even exceed, the number of samples. Here, we show that feature selection with Bayesian inference takes a universal form and reduces to calculating the magnetizations of an Ising model, under some mild conditions. Our results exploit the observation that the evidence takes a universal form for strongly-regularizing priors ---  priors that have a large effect on the posterior probability even in the infinite data limit. We derive explicit expressions for feature selection for generalized linear models, a large class of statistical techniques that include linear and logistic regression. We illustrate the power of our approach by analyzing feature selection in a logistic regression-based classifier trained to distinguish between the letters B and D in the notMNIST dataset.

\end{abstract}

\maketitle

Modern technological advances have fueled the growth of large datasets where thousands, or even millions, of features can be measured simultaneously. Dealing with such a large number of features is often impractical however, especially when the sample size is limited. Feature selection is one promising approach for dealing with such complex datasets \cite{bishop2006pattern, mackay2003information}.  The goal of feature selection is to identify a subset of features that are relevant for statistical tasks such as prediction or classification. Feature selection is a difficult problem because features are often redundant and strongly correlated with each other. Furthermore, the relevance of a feature depends not only on the dataset being analyzed, but also the statistical techniques being employed. While powerful methods for feature selection exist for a handful of techniques such as linear regression \cite{tibshirani1996regression,zou2005regularization}, there is no unified framework for performing feature selection in a computationally tractable manner.  To address this shortcoming, we present a unified approach for Bayesian feature selection that is applicable to a large class of commonly used statistical models.
 
In principle, Bayesian inference provides a unified framework for feature selection \cite{o2004bayesian,mackay2003information}. Unfortunately, Bayesian methods often require extensive Monte Carlo simulations that become computationally intractable for very high-dimensional problems. In the Bayesian framework, a statistical model is defined by its likelihood function, $p_x(y|\boldsymbol{\theta})$, which describes the observed data, $y$, as a function of some features, $\boldsymbol{x}$, and model parameters, $\boldsymbol{\theta}$. This likelihood function is supplemented by a prior, $p(\boldsymbol{\theta})$, that encodes our belief about the model parameters in the absence of data. Using Bayes rule, one can define the posterior probability, $p_x(\boldsymbol{\theta}|y) \propto p_x(y|\boldsymbol{\theta}) p(\boldsymbol{\theta})$, that describes our belief about the parameter $\boldsymbol{\theta}$ given the observed data $y$. Notice, that for a flat (constant) prior, maximizing the posterior distribution is equivalent to maximizing the likelihood and Bayesian inference reduces to the usual maximum likelihood framework. More generally, the log of the prior can be interpreted as a penalty function that   ``regularizes'' the model parameters. For example, the choice of a Gaussian and Laplace prior correspond to imposing a $L2$ and $L1$ penalty on model parameters, respectively \cite{bishop2006pattern}.

Since the ultimate goal of feature selection is to identify a subset of features, it is helpful to introduce a set of indicator variables, $\boldsymbol{s}$, that indicates whether a feature is included in the statistical model, with $s_i=1$ if feature $i$ is included and $s_i=-1$ if it is not. Assuming that we have no \emph{a priori} information about which variables are relevant, the posterior distribution for $\boldsymbol{s}$ can be computed as \cite{mackay1992bayesian}:
\be
p_x(\boldsymbol{s}|y) \propto p_x(y|\boldsymbol{s}).
\label{posterior}
\ee
where the `evidence' is given by:
\be
p_x(y|\boldsymbol{s})= \int d \boldsymbol{\theta}\, p_x (y|\boldsymbol{\theta}, \boldsymbol{s}) p(\boldsymbol{\theta}|\boldsymbol{s})
\label{evidence}
\ee
Feature selection can be performed by choosing features with large posterior probabilities. 

Our central result is that the logarithm of the evidence maps to the energy of an Ising model for a large class of priors we term ``strongly-regularizing priors''. Thus, computing the marginal posterior probabilities for the relevance of each feature reduces to computing the magnetizations of an Ising model. The key property of strongly-regularizing priors is that they affect the posterior probability even when the sample size goes to infinity \cite{fisher2014fast}. This should be contrasted with the usual assumption of Bayesian inference the effect of the prior on posterior probabilities diminishes inversely with the number of samples \cite{mackay2003information}. Strongly-regularizing priors are especially useful for feature selection problems where the number of potential features is greater than, or comparable to, the number of data points. Surprisingly, our results do not depend on the specific choice of prior or likelihood function, under some mild conditions, suggesting that Bayesian feature selection has universal properties for strongly regularizing priors.

Practically, our Bayesian Ising Approximation (BIA) provides an efficient method for feature selection with many commonly employed statistical procedures such as Generalized Linear Models (GLMs). We envision using the BIA as part of a two-stage procedure where the BIA is applied to rapidly screen irrelevant variables, i.e. those features that have low rank in posterior probability, before applying a more computationally intensive cross validation procedure to infer the model parameters with the remaining features.

\section{General Formulas}

For concreteness, consider a series of $n$ independent observations of some dependent variable $y$, i.e. $\boldsymbol{y} = (y_1, \ldots, y_n)$, that we want to explain using a set of $p$ potential features $\boldsymbol{x}=(x_1,\ldots , x_p)$. Furthermore, assume that to each of these potential features, $x_i$, we associate a model parameter $\theta_i$. Let $S$ denote an index set specifying the positions $j$ for which $s_j = +1$. The cardinality of $S$, i.e.\ the number of indicator variables with $s_j = +1$, is denoted $M$. Also, we define a vector $\boldsymbol{\hat{\theta}}$ of length $M$ that contains the model parameters corresponding to the active features, $S$. With these definitions, we define the log-likelihood for the observed data as a function of the active features as ${\cal L}_{X}(\boldsymbol{y}| \boldsymbol{\hat{\theta}})=-\log{p_{X}(\boldsymbol{y}|\hat{\boldsymbol{\theta}})}$. Throughout, we assume that the log-likelihood is twice differentiable. 

In addition to the log-likelihood, we need to specify prior distributions for the parameters. Here, we will work with factorized priors of the form
\be
P(\hat{\boldsymbol{\theta}} ) = \prod_{j \in S} \frac{1}{Z(\l)} e^{-\l f(\hat{\theta}_j) },
\ee
where $\lambda$ is a parameter that controls the strength of the regularization, $f(\theta_j)$ is a twice differentiable convex function minimized at the point $\bar{\theta}_j$ with $f(\bar{\theta}_j) = 0$, and $Z(\l)$ is the normalization constant. As the function is convex, we are assuming that the second derivative evaluated at $\bar{\theta}_j$, denoted ${\del^2f}$, is positive. Many commonly used priors take this form including Gaussian and hyperbolic priors. Plugging these expressions into (\ref{evidence}), yields an expression for the evidence when only the subset of features, $S$, are included:
\be
p_{X}(\boldsymbol{y}|S) = Z^{-M} (\l) \int d \boldsymbol{\hat{\theta}} e^{ \mathcal{L}_{X}(\boldsymbol{y}|\boldsymbol{\hat{\theta}}) - \l \sum_{j \in S} f(\hat{\theta}_j) }
\ee

By definition, this integral is dominated by the second term for strongly regularizing priors. Since the log-likelihood is extensive in the number of data points, $n$, this generally requires that the regularization strength be much larger than the number of data points, $\lambda  \gg n$. For such strongly regularizing priors we can perform a saddle-point approximation for the evidence. This yields, up to an irrelevant constant,
\begin{align}
& {\cal L}_{X}(\boldsymbol{y} | S) = \log{p_{X}(\boldsymbol{y}|S)} \nonumber \\
&= -{1 \over 2}\log{ |I - \Lambda^{-1} H |} + {\Lambda^{-1} \over 2} \boldsymbol{b}'(I -  \Lambda^{-1} H)^{-1} \boldsymbol{b}
\label{evidence2}
\end{align}
where $\Lambda = \l (\del^2 f)$ is a ``renormalized'' regularization strength, 
\be
\boldsymbol{b} = \nabla{ \cal L}_{X}(\boldsymbol{y}|\boldsymbol{\hat{ \theta}})
\label{defb}
\ee
is the gradient of the log-likelihood,  and 
\be
H_{ij}= {\partial^2{ \cal L}_{X}(\boldsymbol{y}|\boldsymbol{ \hat{\theta}}) \over \partial \hat{\theta}_i \partial \hat{\theta}_j}
\label{defH}
\ee
is the Hessian of the log-likelihood, with all derivatives evaluated at  $\hat{\theta}_j=\bar{\theta}_j$. We emphasize that the large parameter in our saddle-point approximation is the regularization strength $\Lambda \gg n$. This differs from previous statistical physics approaches to Bayesian feature selection that commonly assume that  $n\gg \Lambda$ and hence perform a saddle point in $n$ \cite{kinney2014parametric, balasubramanian1997statistical,nemenman2002occam}. The fundamental reason for this difference is that we work in the strongly-regularizing regime where the prior is assumed to be important even in the infinite data limit.

Since for strongly-regularizing priors, $\Lambda$ is the largest scale in the problem, we can expand the log in (\ref{evidence2}) in a power series expansion in $\epsilon = \Lambda^{-1}$  to order $O({\e^3})$ to get: 
\be
{\cal L}_{X}(\boldsymbol{y} | S) \simeq {\e \over 2} ( \text{Tr}[H]  + \boldsymbol{b}' \boldsymbol{b})  + { \e^2 
\over 2} ( {1 \over 2} \text{Tr}[H^2] +  \boldsymbol{b}' H \boldsymbol{b})
\label{LL1}
\ee
One of the most striking aspects of this expression is that it is independent of the detailed form of the prior function, $f(\theta)$. All that is required is that the prior is a twice differential convex function with a global minimum at $f(\bar{\theta}_j) = 0$. Different choices of prior simply ``renormalize'' the effective regularization strength $\Lambda$. 

Rewriting this expression in terms of the spin variables, $\boldsymbol{s}$, we see that the evidence takes the Ising form with:
\be
{\cal L}_{X}(\boldsymbol{y} | \boldsymbol{s})
= \epsilon \left( \sum_i  s_i h_i + \sum_{ij} J_{ij} s_i s_j  \right)
\ee
and couplings given by:
\bea
h_i &=&  {1 \over 4} ( H_{ii} +  b_i^2)  + 2 \sum_j J_{ij}  \label{h}\\
J_{ij} &=&  {\epsilon \over 16} ( H_{ij}^2 + 2 b_i H_{ij} b_j)
\label{couplings}
\eea
Notice that the couplings, which are proportional to the small parameter $\e$, are weak. According to Bayes rule,  ${\cal L}(\boldsymbol{y}| \boldsymbol{s}) = {\cal L}( \boldsymbol{s} | \boldsymbol{y}) + constant$, so the posterior probability of a set of features $\boldsymbol{s}$ is described by an Ising model of the form:
\be
p( \boldsymbol{s} | \boldsymbol{y}) = \mathcal{Z}^{-1} e^{ \sum_i  s_i  h_i + \sum_{ij} s_i s_j J_{ij}}
\label{IsingProb}
\ee
where $\mathcal{Z}$ is the partition function that normalizes the probability distribution. We term this the Bayesian Ising Approximation (BIA) \cite{fisher2014fast}. Finally, for future use, it is useful to define a scale $\L^*$ for which  the Ising approximation breaks down. This scale $\L^*$ can be computed by requiring the that the power series expansion used to derive (\ref{LL1}) converge \cite{fisher2014fast}.

\begin{figure}
\includegraphics[width=8cm]{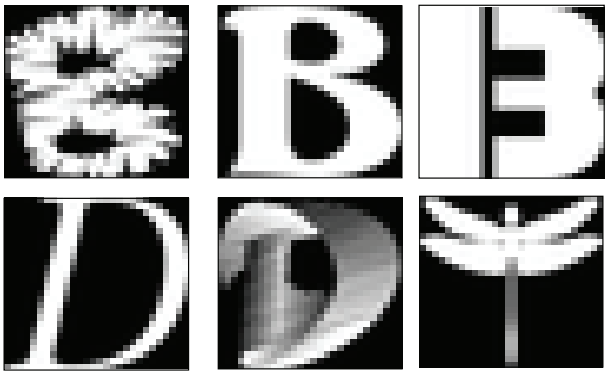}
\caption{{\bf Three randomly chosen examples of Bs (top) and Ds (bottom) from the notMNIST dataset.} Each letter in the notMNIST dataset
is represented by a $28$ by $28$ pixel grayscale image.}
\label{Examples}
\end{figure}

We demonstrated the utility of the BIA for feature selection in the specific context of linear regression in a recent paper, and we can import that machinery here \cite{fisher2014fast}.  It is useful to explicitly indicate the dependence of various expression on the regularization strength $\Lambda$.  We want to compute the marginal posterior probability that a feature, $j$, is relevant: 
\be
p_{\L}(s_j = 1 | \bold{y}) \simeq (1+ m_j(\L) )/2
\label{posterior}
\ee
where we have defined the magnetizations $m_j(\L) = \< s_j \>$. While there are many techniques for calculating the magnetizations of an Ising model, we focus on the mean field approximation which leads to a self-consistent equation \cite{opper20012}:
\begin{align}
m_i(\L) = \tanh \left[  \frac{n^2}{4 \L} \left( h_i(\L) + \HF \sum_{j\neq i}  J_{ij}(\L) m_{j}(\L) \right)  \right]  \nonumber
\end{align}

Our expressions depend on a free parameter ($\L$) that determines the strength of the prior distribution. As it is usually difficult, in practice, to choose a specific value of $\L$ ahead of time it is often helpful to compute the feature selection path; i.e.\ to compute $m_j(\L)$ over a wide range of $\L$'s. Indeed, computing the variable selection path is a common practice when applying other feature selection techniques such as LASSO regression \cite{tibshirani1996regression}. To obtain the mean field variable selection path as a function of $\e = 1/\L$, we notice that $\lim_{\e \to 0} m_j(\e) = 0$ and so define the recursive formula:
\begin{align}
&m_i(\e + \d\e) =			 \nonumber \\
&\tanh \left[  \frac{(\e + \d\e) n^2}{4} \left( h_i\left(\e + \d\e\right) + \HF \sum_{j \neq i} J_{ij}\left(\e + \d\e\right) m_j\left(\e\right) \right)  \right]  \nonumber
\label{recursion}
\end{align}
with a small step size $\d \e \ll 1/\L^*$. We have set $\d \e = 0.05 / \L^{*}$ in all of the examples presented below. 

\begin{figure*}
\includegraphics[width=15cm]{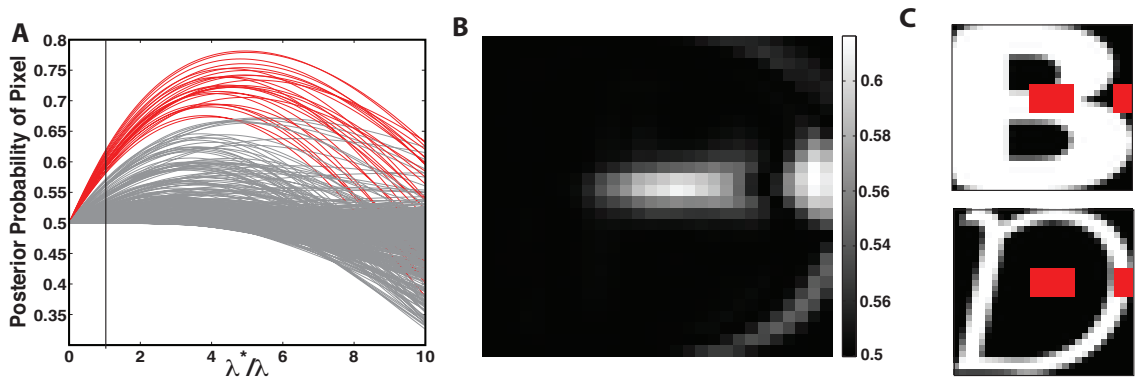}
\caption{{\bf Identifying the most informative pixels for distinguishing Bs and Ds}. {\bf A} We used the Bayesian Ising Approximation (BIA) to perform feature selection for a logistic
regression based classifier designed to classify Bs and Ds. The model was trained 500 using randomly chosen examples of each letter from the notMNIST dataset. The resulting 
feature selection path is shown as a function of the regularization strength, $\lambda$. Each line corresponds to one of 784 pixels. We expect the BIA to breakdown when $\lambda < \lambda^*$. Pixels in red have the highest posterior probability and hence can be used for distinguishing between Bs and Ds. {\bf B} Posterior probabilities of all 784 pixels when $\L/\L^*=1$. {\bf C} Visualization of the most informative pixels labeled in red in part A. }
\label{BIAfigure}
\end{figure*}

\section{Examples}

Logistic regression is a commonly used statistical method for modeling categorical data \cite{bishop2006pattern}. To simplify notation, it is useful also useful define an extra feature variable $x_0=1$ which is always equal to $1$ and a $p+1$-dimensional vector of feature, $\boldsymbol{x}= (x_0, x_1, \ldots, x_{p})$ with corresponding parameters, $\boldsymbol{\theta}=(\theta_0, \theta_1, \dots, \theta_p)$.  In terms of these vectors, the likelihood function for logistic regression takes the compact form
\be
P_{\boldsymbol{x}}(y=1|\boldsymbol{\theta})= 1-P_{\boldsymbol{x}}(y=1|\boldsymbol{\theta})={e^{\boldsymbol{\theta}' \boldsymbol{x}} \over 1+e^{\boldsymbol{\theta}' \boldsymbol{x}}},
\ee
where $\boldsymbol{\theta}'$ is the transpose of $\boldsymbol{\theta}$.
If we have $n$ independent observations of the data (labeled by index $l=1\ldots n$), the log-likelihood can be written as:
\be
{\cal L}_{X}( \boldsymbol{y} |\boldsymbol{\theta})=\sum_l y_l (\boldsymbol{\theta}' \boldsymbol{x}^l)-\log{(1+e^{\boldsymbol{\theta}' \boldsymbol{x}^l})}.
\ee
We supplement this likelihood function with an  $L^2$ norm on the parameters of the form:
\be
p(\boldsymbol{\theta} ) = \prod_{j=1}^p \sqrt{\frac{\l}{2 \pi }} e^{-\l (\theta_j - \bar{\theta}_j)^2/2 }
\ee
with $\bar{\boldsymbol{\theta}} = (\bar{\theta}_0, 0, \ldots, 0)$ and where $\bar{\theta}_0$ is chosen to match the observed probability that $y=1$ in the data,
\be
P_{obs}(y).={e^{\bar{\theta}_0} \over 1+e^{\bar{\theta}_0}}
\ee

Using these expressions, we can calculate the gradient and the Hessian of the log-likelihood:
\be
b_i \equiv {\partial {\cal L}_{X}(\boldsymbol{y}|\boldsymbol{\theta}) \over \partial \theta_i}|_{\bar{\boldsymbol{\theta}}}=\sum_l x_i^l \left( y^l- P_{obs}(y) \right)
\ee
and
\be
H_{ij} = -  \left( P_{obs}(y) (1-P_{obs}(y) \right).\sum_l x_i^l x_j^l
\ee
Plugging in these into (\ref{couplings}) yields the couplings for the Ising model in (\ref{IsingProb}). 

Notice that the gradient, $b_i$, is proportional to the correlations between the $y$ and $x_i$. Furthermore, except for a multiplicative constant reflecting the variance of $y$, $H_{ij}$ is just correlation between $x_i$ and $x_j$. Thus, as in linear regression \cite{fisher2014fast}, the coefficients of the Ising model are related to the correlations between variables and/or the data. In fact, it is easy to show this is the case for all Generalized Linear Models (see Appendix).

To illustrate our approach, we used the BIA for a logistic regression-based classifier designed to classify Bs and Ds in the notMNIST dataset. The notMNIST dataset consists of diverse images of the letters A-J composed from publicly available computer fonts. Each letter is represented as a 28 by 28 grayscale image where pixel intensities vary between 0 and 255.  Figure \ref{Examples} shows three randomly chosen examples of the letters B and D from the notMNIST dataset.  The BIA was performed using 500 randomly chosen examples each letter. Notice that the number of examples (500 or each letter) is comparable to the number of pixels (784) in an image, suggesting that strongly-regularizing the problem is appropriate.

Using the expressions above, we  calculated the couplings for the Ising model describing our logistic-regression based classifier and calculated the feature selection path as a function of ${\L / \L^*}$ using the mean field approximation. As in linear regression, we used $\L^*=n(1+p r)$, where $n=1000$ is the number of samples, $p=784$ is the number of potential features, and $r$ is the root mean squared correlation between pixels (see Figure \ref{BIAfigure}A). Figure \ref{BIAfigure}B shows the posterior probability of all 784 pixels when $\L=\L^*$. To better visualize this, we have labeled the pixels with the highest posterior probabilities in red in the feature selection path in Fig  \ref{BIAfigure}A and in the sample images shown Figure \ref{BIAfigure}C. The results agree well with our basic intuition about which pixels are important for distinguishing the letters B and D.

\subsection{Conclusion}

In conclusion, we have presented a general framework for Bayesian Feature Selection in a large class of statistical models.  In contrast to previous Bayesian approaches that assume that the effect of the prior vanishes inversely with the amount of data, we have used strongly-regularizing priors that have large effects on the posterior distribution even in the infinite data limit. We have shown that in the strongly-regularized limit, Bayesian feature selection takes the universal form of an Ising model. Thus, the marginal posterior probabilities that of each feature is relevant can be efficiently computed using a mean-field approximation. Furthermore, for Generalized Linear Models we have shown the coefficients of the Ising model can be calculated directly from correlations between the data and features. 

Surprisingly, aside from some mild regularity conditions, our approach is independent of the choice of prior or likelihood. This suggests that it maybe possible to obtain general results about strongly-regularizing priors. It will be interesting to further explore Bayesian inference in this new limit. Our approach also gives a practical algorithm for quickly performing feature selection in many commonly employed statistical and machine learning approaches  The methods outlined here are especially well suited for modern data sets where the number of potential features can vastly exceed the number of independent samples. We envision using the Bayesian feature selection algorithm outlined here as part of a two stage procedure. One can use the BIA  to rapidly screen irrelevant variables and reduce the complexity of the dataset before applying a more comprehensive cross-validation procedure. More generally, it will be interesting to further develop statistical physics based approaches for the analysis of complex data. 

\appendix
\section{Appedix: Formulas for Generalized Linear Models}

The gradient. $\boldsymbol{b}$, and Hessian, $H$, of the log-likelihood have particularly simple definitions for Generalized Linear Models (GLMs) which extend the exponential
family of distributions. In the exponential family, we can write the distribution in the form
\be
p(y|\theta)= g(y)e^{ \boldsymbol{\eta}(\theta) \boldsymbol{T}(y) +F(\eta)}
\ee
where $\boldsymbol{T}(y)$ is a vector of sufficient statistics, and the $\boldsymbol{\eta}$ are called natural parameters. Notice, that for these distributions, we have that
\be
{\partial F \over \partial \eta_i} = - \<T_i(y)\>
\ee
and
\be
{\partial^2 F \over \partial \eta_i \partial \eta_j}=  {\rm Cov}[T_i(y), T_j(y)]
\ee
where Cov denotes the covariance (connected correlation function).

In a GLM, we restrict ourselves to distribution to scalar quantities where $T(y)=y$ and say that $\eta = \boldsymbol{\theta} ' \boldsymbol{x}$. Then, we can write the likelihood as
\be
p(y|\boldsymbol{\theta}, \boldsymbol{x})= g(y)e^{ (\boldsymbol{\theta}' \boldsymbol{x}) y+F(\boldsymbol{\theta}'  \boldsymbol{x})}
\ee
If we have $n$ independent data points with $l=1 \ldots n$ then we can write the log-likelihood for such a distribution as
\be
{\cal L}_{X}(\boldsymbol{y}|\boldsymbol{\theta}) = \sum_l \log{g(y^l)} + ( \boldsymbol{\theta} ' \boldsymbol{x}^l) y^l+F(\boldsymbol{\theta}' \boldsymbol{x}^l)
\ee
Using the expressions above for the exponential family and (\ref{defb}) we have that 
\be
b_i = {\partial { \cal L}_{X}(\boldsymbol{y}|\boldsymbol{ \theta}) \over \partial \theta_i} |_{\boldsymbol{\theta} =\bar{\boldsymbol{\theta}}} = \sum_l  y^l x_i^l - x_i^l \<y\>_{\bar{\boldsymbol{\theta}}} 
\ee
where $ \<y\>_{\bar{\boldsymbol{\theta}}}$ is the expectation value of $y$ for choice of parameter $\boldsymbol{\theta} =\bar{\boldsymbol{\theta}}$. If we choose $\bar{\theta}$ to reproduce the empirical probability we get:
\be
b_i  \approx n {\rm Cov} [y,x_i]
\ee
Moreover, the entries of the Hessian are given by:
\be
H_{ij}= {\partial^2{ \cal L}_{X}(\boldsymbol{y}|\boldsymbol{\theta}) \over \partial \theta_i \partial \theta_j}= -\sum_l  x_i^l x_j^l  {\rm Var}[y]_{\boldsymbol{\theta} =\bar{\boldsymbol{\theta}}}
\ee
If we consider standardized variables, $\boldsymbol{x}$, then we can write:
\be
H_{ij} \approx  - n {\rm Cov}[x_i, x_j] {\rm Var}[y]
\ee

\bibliography{refsmain} 

\begin{thebibliography}{11}%
\makeatletter
\providecommand \@ifxundefined [1]{%
 \@ifx{#1\undefined}
}%
\providecommand \@ifnum [1]{%
 \ifnum #1\expandafter \@firstoftwo
 \else \expandafter \@secondoftwo
 \fi
}%
\providecommand \@ifx [1]{%
 \ifx #1\expandafter \@firstoftwo
 \else \expandafter \@secondoftwo
 \fi
}%
\providecommand \natexlab [1]{#1}%
\providecommand \enquote  [1]{``#1''}%
\providecommand \bibnamefont  [1]{#1}%
\providecommand \bibfnamefont [1]{#1}%
\providecommand \citenamefont [1]{#1}%
\providecommand \href@noop [0]{\@secondoftwo}%
\providecommand \href [0]{\begingroup \@sanitize@url \@href}%
\providecommand \@href[1]{\@@startlink{#1}\@@href}%
\providecommand \@@href[1]{\endgroup#1\@@endlink}%
\providecommand \@sanitize@url [0]{\catcode `\\12\catcode `\$12\catcode
  `\&12\catcode `\#12\catcode `\^12\catcode `\_12\catcode `\%12\relax}%
\providecommand \@@startlink[1]{}%
\providecommand \@@endlink[0]{}%
\providecommand \url  [0]{\begingroup\@sanitize@url \@url }%
\providecommand \@url [1]{\endgroup\@href {#1}{\urlprefix }}%
\providecommand \urlprefix  [0]{URL }%
\providecommand \Eprint [0]{\href }%
\providecommand \doibase [0]{http://dx.doi.org/}%
\providecommand \selectlanguage [0]{\@gobble}%
\providecommand \bibinfo  [0]{\@secondoftwo}%
\providecommand \bibfield  [0]{\@secondoftwo}%
\providecommand \translation [1]{[#1]}%
\providecommand \BibitemOpen [0]{}%
\providecommand \bibitemStop [0]{}%
\providecommand \bibitemNoStop [0]{.\EOS\space}%
\providecommand \EOS [0]{\spacefactor3000\relax}%
\providecommand \BibitemShut  [1]{\csname bibitem#1\endcsname}%
\let\auto@bib@innerbib\@empty
\bibitem [{\citenamefont {Bishop}\ \emph {et~al.}(2006)\citenamefont {Bishop}
  \emph {et~al.}}]{bishop2006pattern}%
  \BibitemOpen
  \bibfield  {author} {\bibinfo {author} {\bibfnamefont {C.~M.}\ \bibnamefont
  {Bishop}} \emph {et~al.},\ }\href@noop {} {\emph {\bibinfo {title} {Pattern
  recognition and machine learning}}},\ Vol.~\bibinfo {volume} {1}\ (\bibinfo
  {publisher} {springer New York},\ \bibinfo {year} {2006})\BibitemShut
  {NoStop}%
\bibitem [{\citenamefont {Mackay}(2003)}]{mackay2003information}%
  \BibitemOpen
  \bibfield  {author} {\bibinfo {author} {\bibfnamefont {D.~J.}\ \bibnamefont
  {Mackay}},\ }\href@noop {} {\bibfield  {journal} {\bibinfo  {journal}
  {Information Theory, Inference and Learning Algorithms, by David JC MacKay,
  pp. 640. ISBN 0521642981. Cambridge, UK: Cambridge University Press, October
  2003.}\ }\textbf {\bibinfo {volume} {1}} (\bibinfo {year}
  {2003})}\BibitemShut {NoStop}%
\bibitem [{\citenamefont {Tibshirani}(1996)}]{tibshirani1996regression}%
  \BibitemOpen
  \bibfield  {author} {\bibinfo {author} {\bibfnamefont {R.}~\bibnamefont
  {Tibshirani}},\ }\href@noop {} {\bibfield  {journal} {\bibinfo  {journal}
  {Journal of the Royal Statistical Society. Series B (Methodological)}\ ,\
  \bibinfo {pages} {267}} (\bibinfo {year} {1996})}\BibitemShut {NoStop}%
\bibitem [{\citenamefont {Zou}\ and\ \citenamefont
  {Hastie}(2005)}]{zou2005regularization}%
  \BibitemOpen
  \bibfield  {author} {\bibinfo {author} {\bibfnamefont {H.}~\bibnamefont
  {Zou}}\ and\ \bibinfo {author} {\bibfnamefont {T.}~\bibnamefont {Hastie}},\
  }\href@noop {} {\bibfield  {journal} {\bibinfo  {journal} {Journal of the
  Royal Statistical Society: Series B (Statistical Methodology)}\ }\textbf
  {\bibinfo {volume} {67}},\ \bibinfo {pages} {301} (\bibinfo {year}
  {2005})}\BibitemShut {NoStop}%
\bibitem [{\citenamefont {O'Hagan}\ \emph {et~al.}(2004)\citenamefont
  {O'Hagan}, \citenamefont {Forster},\ and\ \citenamefont
  {Kendall}}]{o2004bayesian}%
  \BibitemOpen
  \bibfield  {author} {\bibinfo {author} {\bibfnamefont {A.}~\bibnamefont
  {O'Hagan}}, \bibinfo {author} {\bibfnamefont {J.}~\bibnamefont {Forster}}, \
  and\ \bibinfo {author} {\bibfnamefont {M.~G.}\ \bibnamefont {Kendall}},\
  }\href@noop {} {\emph {\bibinfo {title} {Bayesian inference}}}\ (\bibinfo
  {publisher} {Arnold London},\ \bibinfo {year} {2004})\BibitemShut {NoStop}%
\bibitem [{\citenamefont {MacKay}(1992)}]{mackay1992bayesian}%
  \BibitemOpen
  \bibfield  {author} {\bibinfo {author} {\bibfnamefont {D.~J.}\ \bibnamefont
  {MacKay}},\ }\href@noop {} {\bibfield  {journal} {\bibinfo  {journal} {Neural
  computation}\ }\textbf {\bibinfo {volume} {4}},\ \bibinfo {pages} {415}
  (\bibinfo {year} {1992})}\BibitemShut {NoStop}%
\bibitem [{\citenamefont {Fisher}\ and\ \citenamefont
  {Mehta}(2014)}]{fisher2014fast}%
  \BibitemOpen
  \bibfield  {author} {\bibinfo {author} {\bibfnamefont {C.~K.}\ \bibnamefont
  {Fisher}}\ and\ \bibinfo {author} {\bibfnamefont {P.}~\bibnamefont {Mehta}},\
  }\href@noop {} {\bibfield  {journal} {\bibinfo  {journal} {arXiv preprint
  arXiv:1407.8187}\ } (\bibinfo {year} {2014})}\BibitemShut {NoStop}%
\bibitem [{\citenamefont {Kinney}\ and\ \citenamefont
  {Atwal}(2014)}]{kinney2014parametric}%
  \BibitemOpen
  \bibfield  {author} {\bibinfo {author} {\bibfnamefont {J.~B.}\ \bibnamefont
  {Kinney}}\ and\ \bibinfo {author} {\bibfnamefont {G.~S.}\ \bibnamefont
  {Atwal}},\ }\href@noop {} {\bibfield  {journal} {\bibinfo  {journal} {Neural
  computation}\ }\textbf {\bibinfo {volume} {26}},\ \bibinfo {pages} {637}
  (\bibinfo {year} {2014})}\BibitemShut {NoStop}%
\bibitem [{\citenamefont
  {Balasubramanian}(1997)}]{balasubramanian1997statistical}%
  \BibitemOpen
  \bibfield  {author} {\bibinfo {author} {\bibfnamefont {V.}~\bibnamefont
  {Balasubramanian}},\ }\href@noop {} {\bibfield  {journal} {\bibinfo
  {journal} {Neural computation}\ }\textbf {\bibinfo {volume} {9}},\ \bibinfo
  {pages} {349} (\bibinfo {year} {1997})}\BibitemShut {NoStop}%
\bibitem [{\citenamefont {Nemenman}\ and\ \citenamefont
  {Bialek}(2002)}]{nemenman2002occam}%
  \BibitemOpen
  \bibfield  {author} {\bibinfo {author} {\bibfnamefont {I.}~\bibnamefont
  {Nemenman}}\ and\ \bibinfo {author} {\bibfnamefont {W.}~\bibnamefont
  {Bialek}},\ }\href@noop {} {\bibfield  {journal} {\bibinfo  {journal}
  {Physical Review E}\ }\textbf {\bibinfo {volume} {65}},\ \bibinfo {pages}
  {026137} (\bibinfo {year} {2002})}\BibitemShut {NoStop}%
\bibitem [{\citenamefont {Opper}\ and\ \citenamefont
  {Winther}(2001)}]{opper20012}%
  \BibitemOpen
  \bibfield  {author} {\bibinfo {author} {\bibfnamefont {M.}~\bibnamefont
  {Opper}}\ and\ \bibinfo {author} {\bibfnamefont {O.}~\bibnamefont
  {Winther}},\ }\href@noop {} {\bibfield  {journal} {\bibinfo  {journal}
  {Advanced mean field methods: theory and practice}\ ,\ \bibinfo {pages} {7}}
  (\bibinfo {year} {2001})}\BibitemShut {NoStop}%
\end{thebibliography}%
\end{document}